\newcommand{\beq} {\begin{eqnarray}}
\newcommand{\eeq} {\end{eqnarray}}
\documentclass[12pt, preprint,preprintnumbers,amsfonts,aps,nofootinbib]{revtex4}
\usepackage{graphicx}
\begin{document}

\title{On the Phenomenology of Strongly Coupled Hidden Sectors}

\author{Nathaniel J. Craig}
\email{ncraig@stanford.edu}
\affiliation{Department of Physics, Stanford University, Stanford, CA 94305-4060} 

\author{Daniel Green}
\email{drgreen@stanford.edu}
\affiliation{SLAC and Department of Physics, Stanford University, Stanford, CA 94305-4060}

\preprint{SU-ITP-09/21}

\begin{abstract}
In models of supersymmetry (SUSY) breaking and mediation, strongly coupled SUSY-breaking sectors can play a significant role in determining the low-energy spectrum of the model.  For example, strong dynamics may provide a natural solution to both the SUSY flavor problem and the $\mu / B \mu$ problem.  Recently, it has been suggested that a large class of these models lead to identical boundary conditions at the SUSY breaking scale. These boundary conditions would severely constrain the models' viability.  We demonstrate that the boundary conditions are instead sensitive to the details of the hidden sector, so that only specific hidden sectors may be ruled out by phenomenological considerations.  We determine the high scale boundary conditions using the operator product expansion of the hidden sector.  The techniques used to determine the beta functions are generally applicable to the RG flow of any approximately conformal hidden sector.  The discrepancy with previously proposed boundary conditions can be traced to the fact that the renormalization group (RG) flow involves multiple fixed points.  
\end{abstract}

\maketitle
\section{Introduction}

Supersymmetry is a attractive scenario for explaining the origin of the weak scale.  Due to tree level sum rules, it is typical to consider models where SUSY is broken in a hidden sector and communicated to the fields of the Minimal Supersymmetric Standard Model (MSSM) through some mediator.  If the SUSY breaking sector is weakly coupled, one often finds that the low energy parameters of the MSSM are determined primarily by the form of mediation rather than the details of the hidden sector.

More recently, it has become clear that the spectrum can be completely altered by hidden sector RG flow when the SUSY breaking sector is strongly coupled.  These models are both of inherent theoretical interest and of use for solving some of the well-known problems in SUSY model building.  One of the best known examples is conformal sequestering \cite{Luty:2001jh,Luty:2001zv}, where large anomalous dimensions of the non-chiral spectrum can suppress dangerous flavor violating effects.  

It is of further interest to understand how strongly coupled hidden sectors influence the parameters of the MSSM in general \cite{Dine:2004dv, Cohen:2006qc}.  It was shown in \cite{Roy:2007nz, Murayama:2007ge} that the hidden sector RG can lead to dynamical solutions to the $\mu$ problem.  In particular, provided only a few conditions on the dynamics of the strongly coupled sector, RG flow may produce $ \mu^2 \simeq B \mu$ from fairly generic initial conditions.

This solution assumes that at some high scale $M$, a K\"{a}hler potential is generated of the form
\beq
 \int d^4 \theta[ (\frac{c_{\mu}}{M} X^{\dag}H_{u} H_{d} + \frac{c_{A_{u,d}}}{M}X^{\dag} H^{\dag}_{u,d} H_{u,d} + h.c.)  + \frac{c_{B\mu}}{M^2}X^{\dag}X H_{u} H_{d} + \frac{c_{m_{u,d}}}{M^2}X^{\dag}X H^{\dag}_{u,d} H_{u,d}],
\eeq
where $X$ is a chiral superfield whose F-term breaks SUSY at the scale $\Lambda \ll M$ ($F_X \sim \Lambda^2$) and $c_{\mu, B\mu,\ldots}$ are the (dimensionful) couplings.  The notation is such that integrating out $F_X$ will generate the terms in the subscript of each $c$ (e.g., $c_\mu$ gives rise to $\mu$).  We will further assume that the $X$ sector is strongly coupled over some range of energies $M>E>\Lambda$ and the scaling dimensions of $X$ and $X^{\dag}X$ ($\Delta_X$ and $\Delta_{X^{\dag}X},$ respectively) satisfy $2\Delta_X < \Delta_{X^{\dag}X}$.  Given these assumptions, from dimensional analysis it is clear that the coupling $c_{B\mu}$ will run to zero faster than $c_{\mu}^2$.  As a result, one finds that $\mu^2 \sim B\mu$ even if $c_{\mu}^2 \ll c_{B\mu}$ at the scale $M$.  

Closer study of the RG flow shows that $c_{\mu}$ and $c_A$ contribute to the running of $c_{B\mu}$ and $c_{m_{u,d}}$ \cite{Roy:2007nz, Murayama:2007ge}.  Therefore, one finds that the Higgs masses $m_{H_{u,d}}$ and $B \mu$ are of the scale $\mu^2$ after RG flow, rather than zero as one might naively expect.  Given that these added contributions involve the strongly coupled fields, it is reasonable to suspect that the exact relation depends on the details of the hidden sector.\footnote{The relation $2\Delta_X < \Delta_{X^{\dag}X}$ does not hold at large N or weak coupling, so by construction one does not have a simple computable example.  Furthermore, these effects appear at order $\lambda^4$, where $\lambda$ is the messenger coupling to $X$.  As a result, the effects of interest are not captured by the results of \cite{Komargodski:2008ax} in the context of general gauge mediation \cite{Meade:2008wd}.}

In several early papers on the subject \cite{Murayama:2007ge, Perez:2008ng}, it was claimed that the boundary conditions at the scale $\Lambda$ are independent of the strongly coupled sector.  Specifically, they argue that the boundary conditions are $m_{H_u}^2 = m_{H_d}^2 = -\mu^2$ and $B\mu =- \mu (A_d + A_u)$.  These boundary conditions were used in \cite{Perez:2008ng,Kim:2009sy,Cho:2008fr} to rule out much of the parameter space of these models. For the most part, the boundary condition $m_{H_{u,d}}^2 = -\mu^2$ makes successful electroweak symmetry breaking rather difficult to achieve.

The above boundary conditions seem quite surprising from the point of view of conformal field theory (CFT).  As discussed in \cite{Craig:2008vs}, we should expect the beta functions for the couplings $c_{B \mu}$ and $c_{m_{u,d}}$ to depend on the coefficients of the operator product expansion (OPE).  In this paper, we calculate the beta functions explicitly and find that the OPE coefficients appear as expected.  When $2\Delta_X < \Delta_{X^{\dag}X}$, we find the boundary conditions at the scale $\Lambda$ take the form
\beq
\label{bc}
c_{m_{u,d}} = \frac{1}{2} C (2+ 2\Delta_X - \Delta_{X^{\dag}X}) (|c_\mu|^2+ |c_{A_{u,d}}|^2) \nonumber\\
c_{B\mu} = \frac{1}{2} C (2+ 2\Delta_X - \Delta_{X^{\dag}X}) Re(c_\mu(c_{A_{u}}^*+c_{A_{d}}^*))
\eeq
where $C$ is a real number that appears in the operator product expansion of $X$ and $X^{\dag}$.  For phenomenological purposes it is important to note that $C$ is in general not an integer, nor is it related to $\Delta_{X}$ or $\Delta_{X^{\dag}X}$.  These boundary conditions can be translated into boundary conditions for $m_{u,d}^2 \sim -c_{m_{u,d}}$ and $B\mu \sim - c_{B\mu},$ keeping in mind $m_{u,d}$ gets additional contributions proportional to $|c_{A}|^2$ from integrating out $F_H$.  However, in order to break SUSY, one must also break conformal invariance.  We will focus on the boundary conditions in terms of $c_{\mu, A, \ldots}$ as they are less sensitive to details of how conformal invariance is broken.

There is a simple way to see that boundary conditions independent of the OPE could not be correct.  At strong coupling, what we mean by ``$X^{\dag}X$" is the lowest dimension non-chiral operator in the OPE of $X$ with $X^{\dag}$.  Let us assume the OPE takes the form 
\beq
X^{\dag}(y) X(0) \sim |y|^{-2 \Delta_X} + |y|^{\Delta_1-2 \Delta_X} c_1 \mathcal{O}_1(0) + |y|^{\Delta_2-2 \Delta_X} c_2 \mathcal{O}_2(0)+\ldots
\eeq
where $c_{1,2}$ are order one coefficients and $\mathcal{O}_{1,2}$ are non-chiral operators of dimension $\Delta_{1,2}$.  In the above discussion $X^{\dag}X \equiv \mathcal{O}_1$.  The claim made in \cite{Murayama:2007ge, Perez:2008ng} is then translated into the statement that $c_{m_{u,d}} - |c_{\mu}|^2-|c_{A_{u,d}}|^2$ runs to zero with scaling dimension $\Delta_1$.  Note that this claim is independent of numerical value of $c_1$.  

However, we can formally consider the limit where $c_1 \to 0$.  In this limit, $\mathcal{O}_1$ has nothing to do with $X^{\dag} X$ so the coupling ($c$) of the operator $c \mathcal{O}_1 H_u H_d$ should run to zero, not to $c_{\mu}^2$.    If the boundary condition were independent of $c_1$, the physics of this decoupling limit would be discontinuous.  This would be a disaster as one would need to distinguish $c_1 =0$ from $c_1 = \epsilon$ for arbitrarily small $\epsilon$ to determine order-one differences in the boundary conditions.

The way in which earlier analyses arrived at the boundary conditions $c_{m_{u,d}} = |c_{\mu}|^2+|c_{A_{u,d}}|^2$ was through a combination of field redefinitions and component analysis made at weak coupling.  We will devote some time to explaining why these techniques fail in some strongly coupled models.  In short, such techniques are good for simplifying RG flows in the vicinity of a single fixed point.  However, when the RG flow involves multiple fixed points, one cannot learn about the global behavior of the flow from a local field redefinition.

The organization is as follows:  In Section II,  we will derive the conditions (\ref{bc}) by directly calculating the beta function using the OPE.  In Section III, we will show the weak coupling analogue of our results and describe how to obtain OPE-like structures at weak coupling.  In Section IV we will discuss in detail why the previous derivations fail to reproduce these results.  We will conclude with a discussion of the phenomenological implications of the boundary conditions derived in Section II.

\section{The OPE and Beta Functions}

In this section, we will compute the beta function for hidden sector renormalization using conformal perturbation theory.  Specifically, we will assume the hidden sector behaves like a superconformal field theory (SCFT) perturbed by irrelevant operators for some range of energies.  In theories where $X$ and $X^{\dag}X$ have anomalous dimensions that are not simply related, the operator product expansion will contain poles and zeros.  These will lead to many features that are not familiar from weakly coupled theories.  We will adopt a renormalization scheme where divergences related to the OPE can be absorbed easily.  Before we perform the calculation, we will briefly explain the reasoning behind our choice of scheme.  Readers familiar with the advantages of Wilsonian schemes for complicated RG flows can skip directly to our calculation in Section \ref{section:beta}.

\subsection{Advantages of Wilsonian Schemes}

The RG scheme we will use is a Wilsonian scheme where integrals over position will be cut off at a scale $a$.  We perform all calculation of the OPEs in superspace, so supersymmetry is explicitly maintained.  The beta functions are determined by requiring that all correlation functions are unchanged when we change the cutoff from $a \to a(1+ \delta l)$.  In order to break SUSY, conformal invariance will also be broken and the RG flow will take us to a weakly coupled fixed point.  Wilsonian schemes of this type are well adapted to cover the RG flow between various fixed points.  

Wilsonian schemes have features which are unfamiliar from the perspective of minimal subtraction schemes like $\overline{MS}$. A Wilsonian scheme involves more than just divergences; all integrals are cut off whether or not they are divergent.  When an integral makes a cutoff dependent contribution to a correlation function, it will contribute to the beta function.  This will lead to two important consequences.  First, operators of any dimension can enter into the beta functions of any other operators.  Of course, to a given level of accuracy the contributions of highly irrelevant operators can be ignored, but in principle all operators that are related by OPEs should be present.  Second, we will find that important contributions to the flow of irrelevant operators arise from terms that vanish as $a \to 0$.

\begin{figure}[t] 
   \centering
   \includegraphics[width=4in]{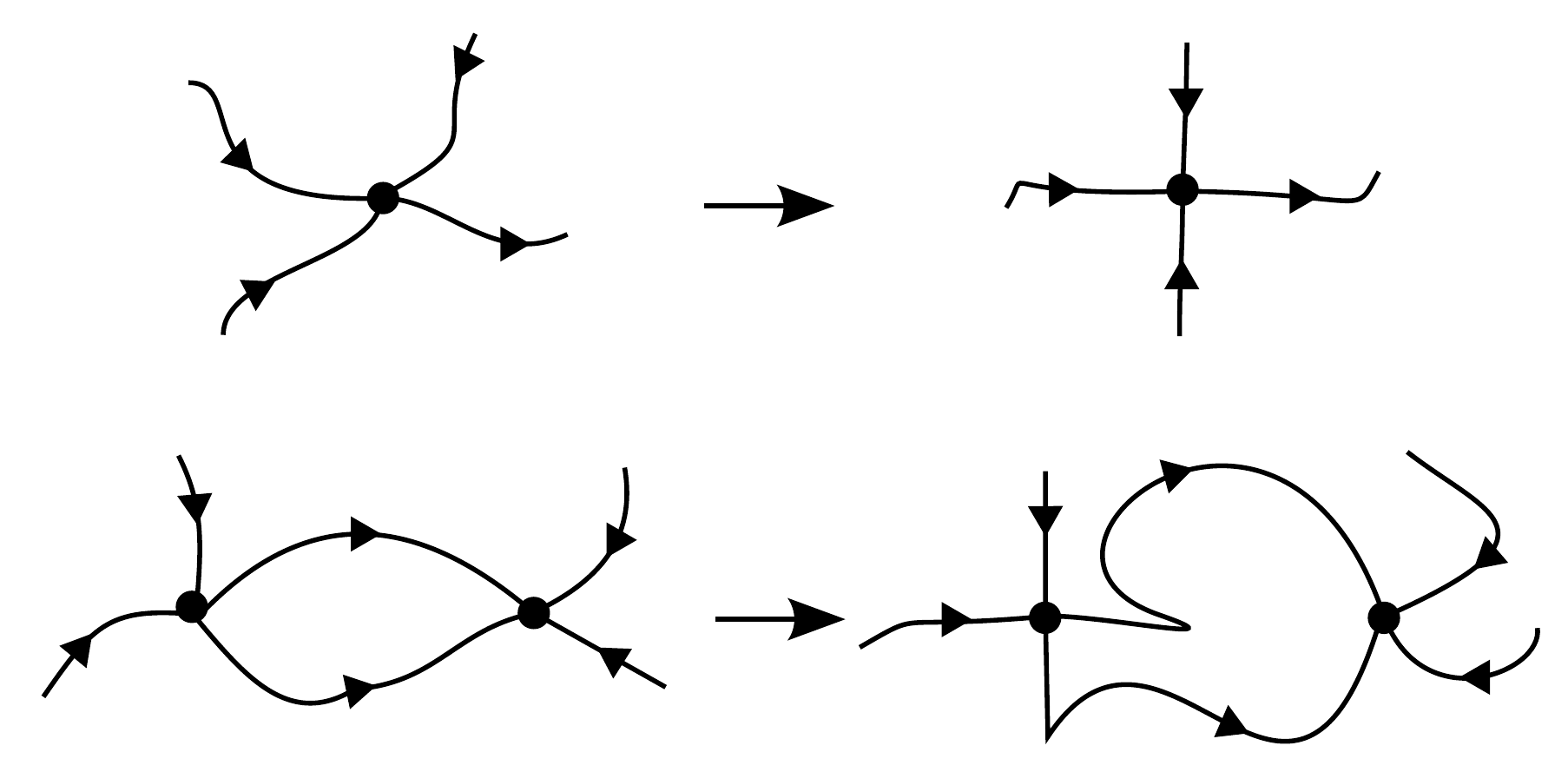} 
      \caption{Linearizing RG flows: single and multiple fixed points.}
   \label{fig:flow}
\end{figure}

If we were only interested in a single fixed point, this choice of scheme could be easily related to a scheme like $\overline{MS}$ \cite{Gaberdiel:2008fn}.  As we will see later, it is always possible to redefine couplings locally to remove any terms in the beta functions that do not arise from logarithmic divergences.  These redefinitions of the couplings are equivalent to field redefinitions that trivialize the RG flow (near a weakly coupled fixed point these are the field redefinitions that appear in \cite{Perez:2008ng}).  However, when flowing between fixed points it should be clear that locally linearizing the RG flow will not simplify the global behavior of the flow.  This can be most easily seen by visually representing the flow in coupling space as in Fig. \ref{fig:flow}.

As one flows between fixed points, both the dimensions of operators and the degree of divergences change.  We should keep track of all operators in the beta functions until it is clear that their contributions are a sub-leading effect.  For example, dangerously irrelevant operators should not be omitted from the beta functions, as they become relevant in the IR despite being irrelevant in the UV.  Wilsonian schemes are better suited for these types of situations as the beta functions are continuous along the flow.

In contrast to many typical scenarios, the choice of scheme is of critical importance to the situation studied here. Typically, one is interested in universal properties of a theory -- aspects of the model that are insensitive to the UV completion.  Contributions from irrelevant operators give `non-universal' contributions in the sense that they depend on the values of irrelevant couplings determined in the UV.  When studying flows towards a fixed point, these contributions become smaller and smaller when flowing to the IR.  However, the Higgs masses we are studying here come directly from irrelevant operators and are sensitive the UV values of the couplings.  Although this is `non-universal', it is the leading effect.  Furthermore, we can embed these models in a UV completion (like a simple model of gauge mediation) in which the values of the couplings in the UV are known. For this reason, a Wilsonian RG scheme is most appropriate for the problem at hand.

\subsection{Beta Functions from OPEs} \label{section:beta}

Conformal perturbation theory is a broadly applicable language that is useful for determining the behavior of correlation functions in the vicinity of any fixed point (not just weakly coupled ones).  For simplicity, we will perform all calculations in Euclidean signature; results in Lorentzian signature may be obtained by analytic continuation.  

In general, one is interested in theories that formally can be described by an action of the form
\beq
S = S_{CFT} + \int d^d x \left( c_1 \mathcal{O}_1 + c_2 \mathcal{O}_2 + \ldots \right),
\eeq
where $\mathcal{O}_i$ are operators in the CFT with scaling dimension $\Delta_i$.  As a simple example, let us begin with a supersymmetric toy model.  Consider perturbing a 4d superconformal theory by 
\beq
S = S_{CFT}+ \int d^4 x \left[ \int d^2\theta( \lambda_1 a^{\Delta_1 -3} \mathcal{O}_1 + h.c.)+ \int d^4 \theta  \lambda_2 a^{\Delta_2 -2} \mathcal{O}_2 \right]
\eeq
where we have introduced explicit dependence on the cutoff $a$ to make the couplings $\lambda_{1,2}$ dimensionless.  Here $\Delta_{1}$ and $\Delta_{2}$ are the conformal dimensions of the lowest component of the chiral superfield $\mathcal{O}_1$ and non-chiral superfield $\mathcal{O}_2,$ respectively.  If we are computing some correlation function at order $|\lambda_1|^2$ we find the contribution
\beq
\label{corr}
\left \langle \ldots |\lambda_1|^2 a^{2\Delta_1 -6} \int d^4 x \int d^4 y \int d^4 \theta \mathcal{O}_1(y) \mathcal{O}^{\dag}_1(x) \right \rangle
\eeq
where $\ldots$ represent any local operators.  Assume now that the OPE takes the form
\beq
\mathcal{O}_1(y) \mathcal{O}^{\dag}_1(x) \sim |\tilde{r}|^{-2 \Delta_1} + c_{112} |\tilde{r}|^{\Delta_2-2 \Delta_1} \mathcal{O}_2(x) + \ldots
\eeq
where $c_{112}$ is a real number and $\tilde{r}^{\mu} = y^{\mu}-x^{\mu} - i \theta_y \sigma^{\mu} \bar{\theta}_x+i\theta_x \sigma^{\mu} \bar{\theta}_y - i (\theta_y - \theta_x) \sigma^{\mu} (\bar{\theta}_y - \bar{\theta}_x)$.  The OPE of an $\mathcal{N}=1$ supersymmetric theory contains many other types of terms \cite{Osborn:1998qu}; however, only the above contributions will play a significant role in the problem at hand.  The coefficient $c_{112}$ can also be determined from the three point function $\langle \mathcal{O}_1 \mathcal{O}^{\dag}_1 \mathcal{O}_2 \rangle \propto c_{112}$.

Using this OPE and defining $r = |y-x|$ while setting $\theta_x = \theta_y$, we can simplify (\ref{corr}) to obtain
\beq
\label{corr2}
\left \langle \ldots c_{112} |\lambda_1|^2 a^{2\Delta_1 -6}  \int d^4 x \int d^4\theta\mathcal{O}_2(x) \int_a^L dr 2 \pi^2 r^3r^{\Delta_2-2 \Delta_1} \right \rangle,
\eeq
where $L$ is an infrared cutoff added to ensure that our integrals are well-defined, but will play no role in our analysis.

To determine the $\beta$ functions,\footnote{Our beta functions are defined by $\beta_g = \frac{\delta g}{\delta l}$.  Note that our beta function is related by a minus sign to the usual high energy conventions since we are flowing to the IR, not the UV.} we wish keep the correlation function fixed while making the change $a \to a(1+ \delta l)$.  The term $|\lambda_1|^2 a^{2\Delta_1 -6}$ results in the leading term in the beta function for $\lambda_1$ of the form $\beta_{\lambda_1} = \lambda_1 (3 - \Delta_1)$.  The change in the cutoff of the integral also introduces a term of the form
\beq
-(\delta l) 2\pi^2 c_{112} |\lambda_{1}|^2 a^{\Delta_2 - 2} \int d^4 x d^4 \theta \mathcal{O}_2(x).
\eeq
This change in the correlation function can be removed by a shift of $\lambda_2,$ which introduces a new contribution to the beta function $\beta_{\lambda_2}$.  Including the contribution from the conformal dimension, we then have
\beq
\beta_{\lambda_2} = \lambda_2 (2 - \Delta_2) - (2 \pi^2 c_{112}) |\lambda_1|^2.  
\eeq

In order to solve the RG flow explicitly, we may linearize the flow by a redefinition of the couplings -- i.e., the RG flow is a system of linear differential equations that may be diagonalized.  Specifically, given a system of differential equations
\beq
\frac{\partial |\lambda_1|^2}{\partial l} &= &2 A |\lambda_1|^2 \nonumber \\
\frac{\partial \lambda_2}{\partial l} &=& B \lambda_2 - C  |\lambda_1|^2,
\eeq
for some constants $A,B,C,$ we may take linear combinations of these equations to obtain
\beq
\frac{\partial{|\lambda_1|^2}}{\partial l} &=& 2 A |\lambda_1|^2 \nonumber \\
\frac{\partial (\lambda_2 - \kappa |\lambda_1|^2)}{\partial l} &=& B (\lambda_2 - \kappa  |\lambda_1|^2),
\eeq
for some number $\kappa$.  Solving for $\kappa,$ we find
\beq
\label{kappa}
\kappa = \frac{C}{B-2A}
\eeq
In the case of our simple toy model, given the beta functions determined above we find that
\beq
\kappa = \frac{2 \pi^2 c_{112}}{2 \Delta_1 - \Delta_2 -4}.
\eeq
It is interesting to note that this procedure does not work when $2 \Delta_1 - \Delta_2 -4 = 0$ (i.e., $2A = B$) which is precisely the case where (\ref{corr2}) is logarithmically divergent.  For the case of a single fixed point, this is the origin of the claim that only logarithmic divergences are universal.

The calculation of the beta functions for the full theory of Higgs fields $H_u, H_d$ coupled to a hidden sector superfield $X$ is essentially identical to the toy model.  The only complication is that we must perform one of the $\int d^4 \theta$ integrals.  Now we are interested in the Lagrangian
\beq
\label{model}
\int d^4 \theta \left[ (\frac{c_{\mu}}{a^{\Delta_X}}X^{\dag} H_{u} H_{d} + \frac{c_{A_{u,d}}}{a^{\Delta_X}}X^{\dag} H^{\dag}_{u,d} H_{u,d} + h.c.)  + \frac{X^{\dag}X}{a^{\Delta_{X^{\dag}X}}}  (c_{B\mu}H_{u} H_{d} +c_{m_{u,d}} H^{\dag}_{u,d} H_{u,d}) \right].
\eeq
We will assume that $X$ is a chiral primary of a SCFT with dimension $\Delta_X > 1$ and that $X^{\dag}X$ is the lowest-dimension primary operator in the OPE of $X$ and $X^{\dag}$.  In the interest of solving the $\mu$ problem, we will also assume that the dimensions of these operators satisfy $\Delta_{X^{\dag}X} > 2\Delta_X > 2$.

Consider again computing correlation functions, now at quadratic order in $c_{\mu}$.  We find the following term:
\beq
\label{term}
|c_{\mu}|^2 a^{2\Delta_{X}}\int d^4 x d^4 \theta_x X^{\dag} H_{u} H_{d}(x)\int d^4 y d^4 \theta_y X H_{u}^{\dag} H_{d}^{\dag}(y).
\eeq
Because $X$ and $H$ are decoupled in the absence of the perturbations, we can use their OPEs independently.  For fields at a free fixed point we find
\beq
H_{u,d}^{\dag}(y) H_{u,d}(x) = (4\pi^2)^{-1}|\tilde{r}|^{-2}+ H^{\dag}_{u,d}H_{u,d}(x),
\eeq
where we have defined the normalization of $H$ to give the factor of $(4 \pi^2)^{-1}$ in order to be consistent with the canonically-normalized weakly-coupled fields.  The OPE for $X$ will be nontrivial, taking the form \cite{Osborn:1998qu}
\beq
X(y) X^{\dag}(x) \sim |\tilde{r}|^{-2 \Delta_X} + C |\tilde{r}|^{\Delta_{X^{\dag}X}-2 \Delta_X} X^{\dag}X(x) + \ldots
\eeq
where $C$ is a real number and $\ldots$ represent higher dimension operators.  For the hidden sector, we have normalized the operators to have unit OPE coefficient for the identity operator.\footnote{This clearly differs from the standard weak coupling value of $(4\pi^2)^{-1}$ but is the commonly used normalization used in CFT.  The important quantity is $C$ whose definition is typically given using this choice of normalization.}  Using the OPEs, one can again simplify the expression (\ref{term}).  We will focus on just one of the terms that is generated, as the analysis for all other possible contributions is identical.

One term that results from the OPE will take the form
\beq
\frac{|c_{\mu}|^2 a^{2\Delta_{X}}}{4 \pi^2}\int d^4 x d^4 \theta_x  X^{\dag}X H_{u}^{\dag} H_{u}(x) \int d^4 y d^4 \theta_y C |\tilde{r}|^{\Delta_{ X^{\dag}X} - 2 \Delta_X-2}.
\eeq
As before, we will redefine $r = |y-x|$.  However, we must now perform the integral over $\theta_y$.  One should do these integrals while maintaining supersymmetry, so the natural choice is a local super-shift $\theta_y \to \theta_y + \theta_x$ and $y \to y + i \theta_y \sigma \bar{\theta}_x - i \theta_x \sigma \bar{\theta}_y$ to remove all the $\theta_x$ dependence.  The $\theta_y$ integral is then equivalent to the Laplacian acting on $r^{\Delta_{X^{\dag}X} - 2 \Delta_X-2}$.  Using $\partial_{\mu}\partial^{\mu} = \partial^2_r + 3r^{-1} \partial_r + r^{-2}\partial_{\Omega}$, we get
\beq
\label{OPE}
\frac{|c_{\mu}|^2 a^{2\Delta_{X}}}{4 \pi^2}\int d^4 x d^4 \theta_x  X^{\dag}X H_{u}^{\dag} H_{u}(x) \int_{a}^{L} dr \frac{C (2 \pi^2)(2\Delta_X - \Delta_{X^{\dag}X})(2+2\Delta_X - \Delta_{X^{\dag}X})}{ r^{2 \Delta_X+4-\Delta_{ X^{\dag}X} }}
\eeq
By the same logic as in the toy model, we will get a contribution to the beta function of $c_{m_u}$ of the form 
\beq
\label{beta}
\beta_{c_{m_u}} = -\Delta_{X^{\dag}X} c_{m_u} - \frac{1}{2}C (2\Delta_X - \Delta_{X^{\dag}X})(2+2\Delta_X - \Delta_{X^{\dag}X}) c_{\mu}^2
\eeq
There will be a similar contribution to the beta function of $c_{m_d}$.  One can check that there will also be contributions to the beta function of $c_{m_{u,d}}$ proportional to $|c_{A_{u,d}}|^2$ and a contribution to $c_{B\mu}$ of order $Re[c_{\mu}(c^{*}_{A_u}+c^*_{A_d})],$ all with the same coefficient as (\ref{beta}).  

Putting these results together with (\ref{kappa}), we find the following combinations of terms run to zero with scaling dimension $\Delta_{X^{\dag}X}$: 
\beq
\label{result}
c_{m_{u,d}} - \frac{1}{2} C (2+2\Delta_X - \Delta_{X^{\dag}X}) (|c_{\mu}|^2 + |c_{A_{u,d}}|^2)  \\
c_{B\mu} - \frac{1}{2} C (2+2\Delta_X - \Delta_{X^{\dag}X}) Re[c_{\mu}( c^*_{A_{u}}+ c^*_{A_d})]
\eeq
These can be related to mass terms after SUSY breaking, keeping in mind that there will be a contribution to the masses $m_{H_{u,d}}^2$ coming from $|c_A|^2$ after integrating out the F-terms of $H_{u,d}$.  

It is worth making a few comments about these results.  The first thing to note is that when $\Delta_{X^{\dag}X} = 2 \Delta_X$ we get no contribution to the beta function!  This is clear in the limit where the hidden sector is free because there is nothing to close the loop diagram. Such behavior will hold more generally even if the hidden sector is strongly interacting.  This is only true to quadratic order in the perturbations away from the CFT, but includes all orders in the CFT sector.

\section{OPE and Weak Coupling}

At first glance, it seems difficult to make contact between our result in (\ref{result}) and intuition gained at weak coupling.  Although we are interested in strongly coupled phenomena, there is nothing about the above result that requires the fixed point to be strongly coupled.  In particular, it is not obvious from a weakly-coupled perspective how to see the appearance of some additional parameter $C$ in the beta functions that is not related to the dimensions of operators.  In this section, we will show how these features appear at weak coupling.

\begin{figure}[t] 
   \centering
   \includegraphics[width=2in]{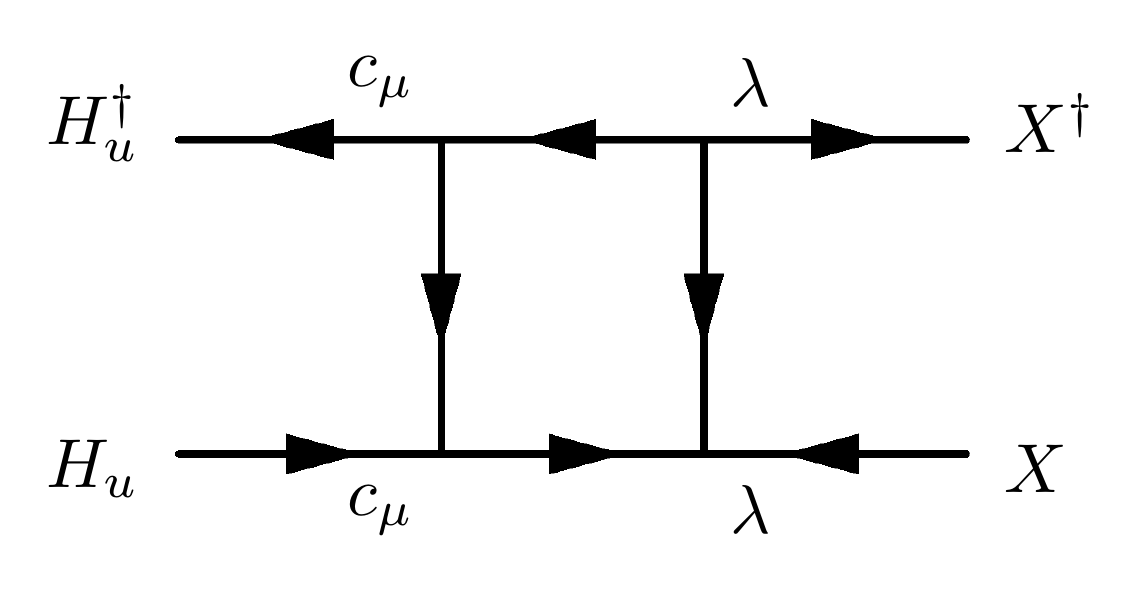} 
    \includegraphics[width=2in]{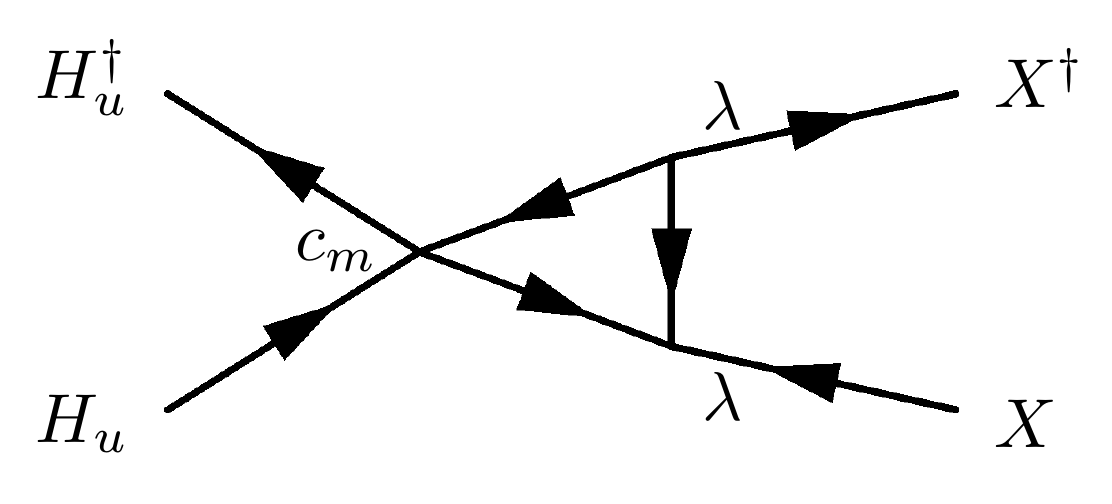} 
   \caption{Diagrams contributing to $\beta_{c_{m_u}}$ at one loop in a weakly coupled $\lambda X^3$ toy SCFT.}
   \label{fig:weak1}
\end{figure}

As a toy model, let us consider the renormalization of (\ref{model}) where $X$ is a weakly coupled field with canonical K\"{a}hler potential and self-interactions described by a superpotential $W = \lambda X^3$ (the choice of $\lambda X^3$ is arbitrary; nothing in this section is particularly sensitive to the specific self-interactions).  In order to reproduce (\ref{beta}), we are interested in computing the anomalous dimension of $c_{m_u}$ from superpotential interactions and the contribution to the beta function $\beta_{c_{m_u}}$ proportional to $|c_{\mu}|^2$.  These diagrams are shown in Fig. \ref{fig:weak1}.

The first confusion is that the diagrams have the same structure -- i.e. at a given order in $\lambda,$ the only difference between the two diagrams are the vertices involving the Higgs; all insertions of the toy SCFT interaction $\lambda$ have the same structure. Diagrammatically, this might lead one to expect $c_{m_u}$ and $|c_\mu|^2$ to appear together in $\beta_{c_{m_u}}$ in the same relative form at all orders in $\lambda.$  Indeed, at one loop both terms in the beta function have the same form.  Even if we include the anomalous dimension of $X$, it would seem that the coefficients of the two terms in the beta function are related.  At weak coupling, the anomalous dimension of $X$ will come from wavefunction renormalization.  As a result, the beta functions will take the form (ignoring all loops involving standard model fields)

\beq
\beta_{c_{\mu}} &=& -(\Delta_X) c_{\mu} \nonumber \\
\beta_{c_m} &=& -(\gamma_{X^{\dag}X} + 2 \Delta_X) c_m + \gamma_{X^{\dag}X} |c_{\mu}|^2 ,
\label{weak}
\eeq
where $\gamma_{X^{\dag}X} = \frac{|\lambda|^2}{16 \pi^2}$ is the anomalous dimension of $X^{\dag}X$ computed by the diagrams in figure Fig. \ref{fig:weak1}. 

We can see that there is precise agreement between (\ref{beta}) and (\ref{weak}) at weak coupling as follows: At weak coupling $\gamma_{X^{\dag}X} \ll 1$, so let us expand to leading order.  Similarly, at $\lambda=0$ the OPE is trivial and $C=1$.  As a result, our result (\ref{beta}) at weak coupling gives gives
\beq
\beta_{c_m} =  -(\Delta_{X^{\dag}X}) c_m + (\Delta_{X^{\dag}X}-2\Delta_X) )|c_{\mu}|^2.
\eeq
where we have used $2 + 2 \Delta_X - \Delta_{X^{\dag}X} \sim 2$.  If we define $\gamma_{X^{\dag}X} = \Delta_{X^{\dag}X}-2\Delta_X$ at weak coupling, we see our result exactly reproduces the ``model independent" boundary condition $m_H^2 \simeq - \mu^2$ ! Thus the OPE calculation gives the known result at weak coupling. However, we also see that as $2 \Delta_X - \Delta_{X^{\dag}X} \sim \mathcal{O}(1)$ we have deviations from this result.  Thus, it should be clear that the ``model independent" result is an accident of weak coupling.

One may also wish to see the origin of $C \neq 1$ from a diagrammatic perspective.  This is far less clear, as it would seem that one always has a trivial OPE inherited from the free fixed point plus perturbative contributions (recall the OPE coefficients appear in the 3 point functions).  In particular, it is not completely obvious that $C$ is independent of the anomalous dimensions.  The key to understanding the origin of the non-trivial OPE is to note that the OPE is the expansion in terms of the primary operators of the CFT.  In particular, it is a sum over all the operators with well-defined conformal dimensions.  We will see that $X^{\dag}X$ is not such an operator at finite $\lambda$ and we will be forced to do a field redefinition if we want to work with operators with well-defined scaling dimensions.

As we mentioned previously, the dimensions of operators change during flows between fixed points.  We are computing the beta functions in a Wilsonian scheme, so we keep track of the contributions from all operators in the theory.  Irrelevant operators may become relevant along the flow so one can only determine which contributions are unimportant after studying the behavior of the full RG flow.

We will apply this logic to the weakly coupled model, with the idea that we may want to eventually take the coupling to be large.  In order to account for all possible operators during RG flow, we should include the renormalization of increasingly irrelevant operators of the form $\int d^4 \theta \alpha_{i} (X^{\dag } X)^i H^{\dag}_{u,d}H_{u,d}$ where $i$  is an integer.\footnote{There are many other operators we should also include, but this subset is sufficient for our present purposes.} Ignoring wavefunction renormalization, let us consider the beta functions for all $\alpha_{i}$.  If we are interested in what happens when $\lambda \sim 1$, we will imagine resumming all the $\lambda$ dependence.

\begin{figure}[t] 
   \centering
   \includegraphics[width=2in]{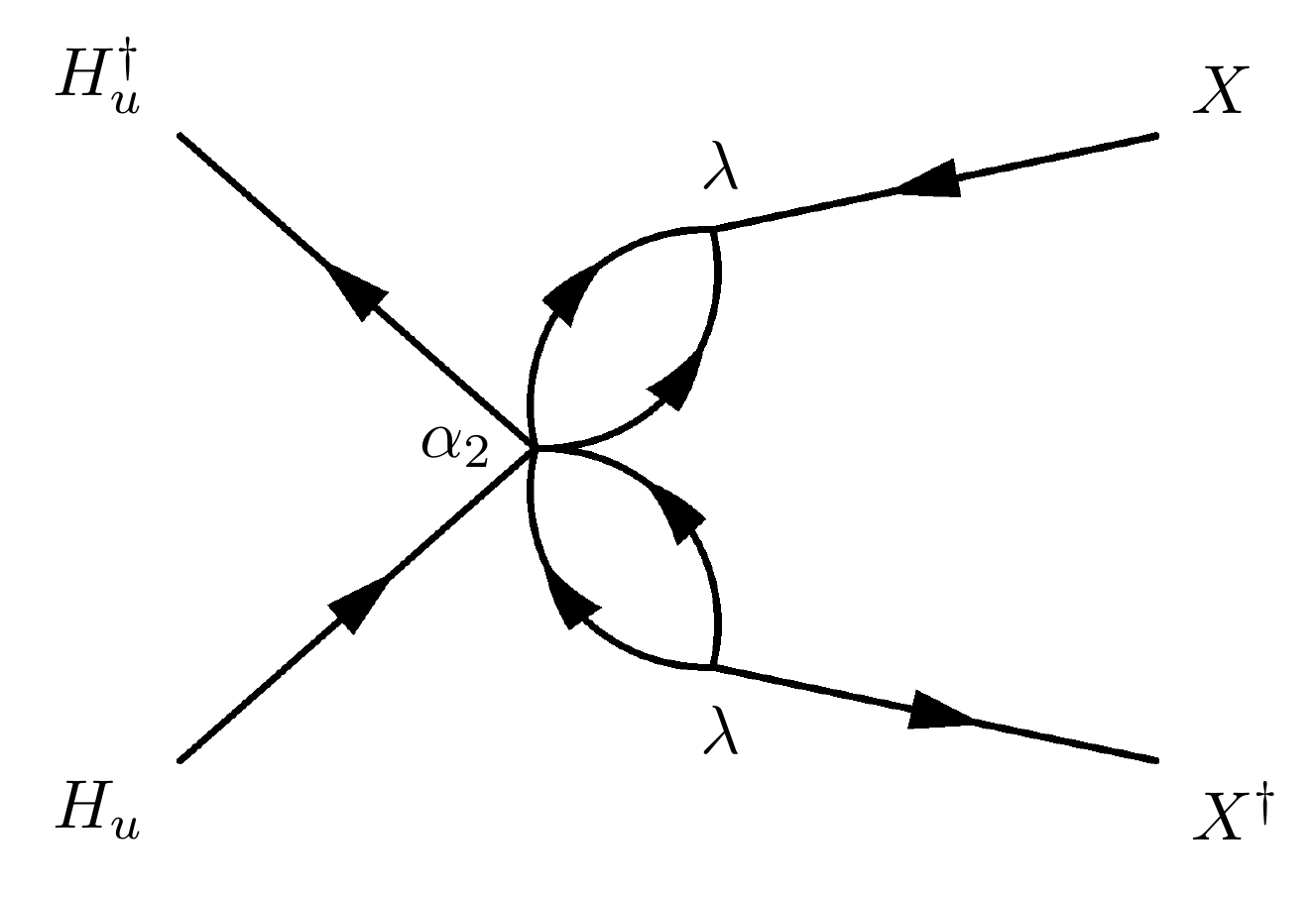} 
    \includegraphics[width=2in]{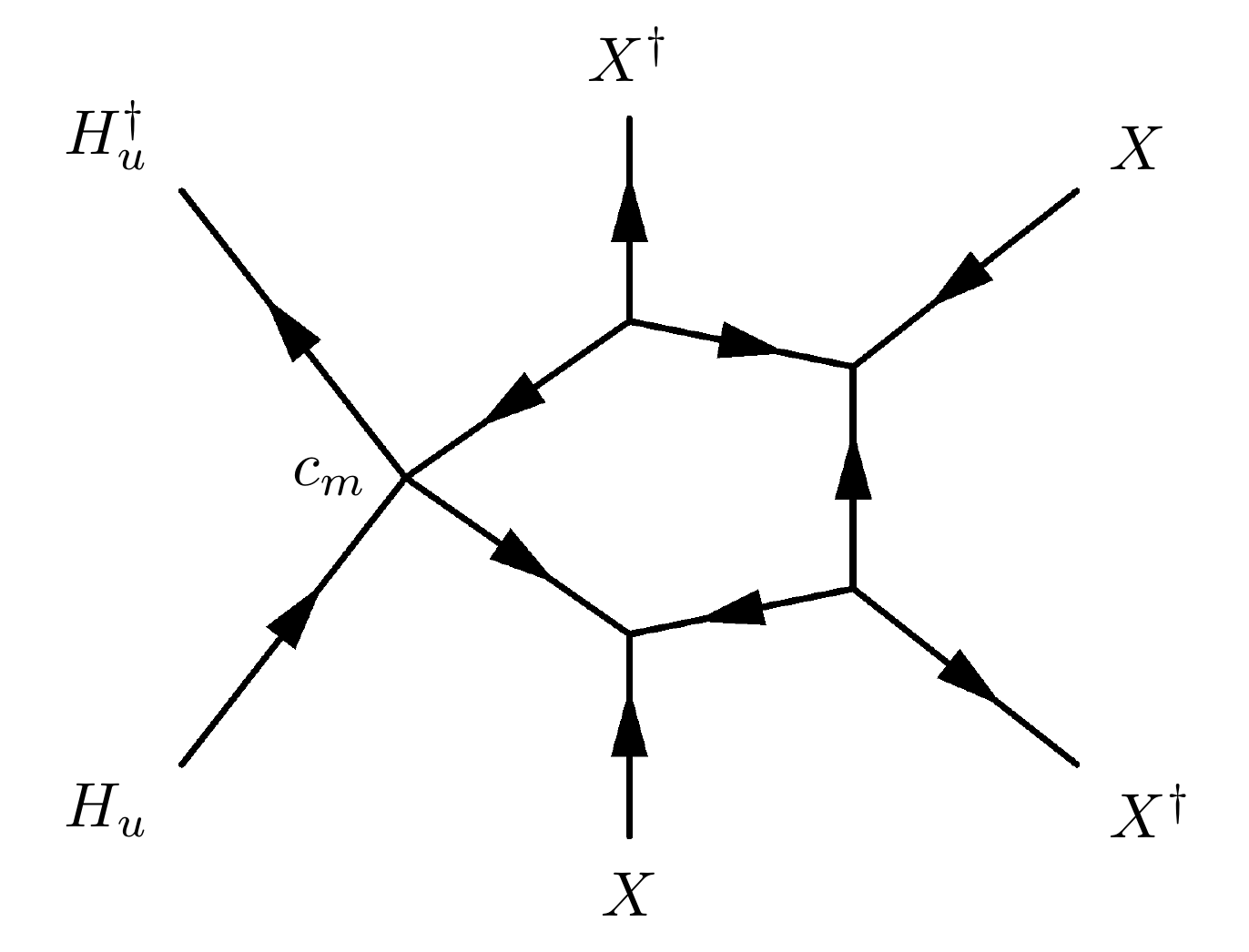} 
   \caption{High-order diagrams contributing to $\beta_{c_{m_u}}$ (left) and $\beta_2$ (right) at one loop in a weakly coupled $\lambda X^3$ toy SCFT.}
   \label{fig:weak2}
\end{figure}

Considering diagrams like those in Fig. \ref{fig:weak2}, it should be clear that the beta functions of $c_{m_{u,d}}$ and the $\alpha_{i}$ take the form
\beq
\beta_{c_{m_{u,d}}} &=&  \gamma_{11} c_{m_{u,d}} + \eta_{1} |c_{\mu}|^2 + \sum_i \gamma_{1i} \alpha_i + \ldots \nonumber \\
\beta_{i} &=& \gamma_{ii} \alpha_{i} +  \gamma_{i1} c_{m_{u,d}} + \eta_{i} |c_{\mu}|^2 + \sum_{j}  \gamma_{ij} \alpha_{j} +\ldots
\eeq
Here the $\eta_i$'s and $\gamma_{ij}$ are some functions of $\lambda$ (possibly starting at high order in $\lambda$).  It is important to note that $\alpha_j$ contributes to the RG flow of $c_{m_{u,d}}$.  Furthermore, by our calculations in section \ref{section:beta}, $\eta$ will differ from $-\gamma_{11}$ and $\eta'$.

In order to apply the analysis of Section II to this model, we should transform to the basis with couplings that diagonalize the matrix of anomalous dimensions.  Specifically, we want to pick out the non-chiral operator that has the smallest anomalous dimension; this is the analogon of a primary operator of the CFT. Therefore we will rotate the basis to be diagonal, where $\tilde{c}_{m_{u,d}} = a_0 c_{m_{u,d}} +\sum_j a_j \alpha_j$ for some real numbers $a_0, a_j$ such that $\tilde{c}$ diagonalizes $\gamma_{ij}$ with eigenvalue $\Delta_{X^{\dag}X}$.

To determine how $c_{\mu}$ influences the running of $\tilde{c},$ we need to know how it contributes to the running of all the $\alpha_j$'s -- not just $c_{m_{u,d}}$.  The terms in the beta functions that appear with $\eta_i$ ensure that in this new basis there will be some arbitrary, order-one factors related to the change of basis; these factors then appear in front of $|c_{\mu}|^2$ in the beta function of $\tilde{c}_{m_{u,d}}$, and should be thought of as the weakly-coupled analogue of the OPE coefficient $C.$  One can think of this change of basis to diagonalize the matrix of anomalous dimensions as the weak coupling analogue of the OPE.

It should also be stressed that this type of behavior is generic in the study of RG flows between fixed points.  In general, one expects that the primary operators at the IR fixed point arise as a linear combination of operators at the UV fixed point.  This is beautifully demonstrated in \cite{Zamolodchikov:1987ti} for RG flows between minimal models in two dimensions. 

\section{Field Redefinitions and Components}

One of the standard arguments for model-independent boundary conditions comes from field redefinitions at a trivial fixed point \cite{Cohen:2006qc}.  This argument assumes that the hidden sector is weakly coupled in the UV and flows to a strongly coupled fixed point in the IR.  Consider, for example, a toy model described by the K\"{a}hler potential
\beq
\label{field}
\int d^4x d^4 \theta \left[ Q^{\dag}Q + (a X Q^{\dag}Q + h.c.) + b X^{\dag}X Q^{\dag}Q \right],
\eeq
where $Q$ is a weakly coupled field, $X$ is our hidden sector field, and $a$ and $b$ are coupling constants.  In the weakly-coupled UV, one can do a field redefinition $\tilde{Q} = (1+aX)Q$ to remove the linear term and thereby arrive at
\beq
\int d^4x d^4 \theta \left[ Q^{\dag}Q + (b-|a|^2) X^{\dag}X Q^{\dag}Q \right].
\eeq
The argument is that $X^{\dag}X$ runs to zero with dimension $\Delta_{X^{\dag}X}$ and hence $b-|a|^2$ also runs to zero.  This would seem to be direct contraction with our results from the OPE, so it merits discussing where such field-redefinition arguments break down

The easiest way to understand the error is to think of the renormalization group geometrically \cite{Wilson:1973jj,Zamolodchikov:1987ti,Kutasov:1988xb,Lassig:1989tc} as suggested by Fig. \ref{fig:flow}.  The RG flow of the system acts on the space of couplings and the operators form the tangent space.  We are free to pick coordinates on this space, which corresponds to defining the operators and couplings.  In particular, we can always choose the coordinates to be locally flat at some point of interest.  However, if the space is not globally flat, one cannot trivialize the RG flow globally.

Since the field redefinition is just a diffeomorphism on the space of couplings, one would expect that the RG flows should be related by the same map.  However, the field redefinition above does not seem to give the result from section \ref{section:beta} mapped by $b \to b - |a|^2$.  The mistake being made is to assume that the scheme for calculating the beta functions is unchanged by the redefinition.  In fact, for the physics in the two bases to be related, one must change the scheme \cite{Latorre:2000qc, Einhorn:2001kj}.  One can see this by considering the S-matrix.  The field redefinition should leave the S-matrix invariant.  But the S-matrix also obeys the Callan-Symanzik equation 
\beq
\left(a \partial_a - \sum_g  \beta_g \partial_g \right) S(g, a) = 0
\eeq
where $a$ is the cutoff, $g$ is the set of couplings and $S(g,a)$ is any S-matrix element.  Assume for some value of the cutoff we can perform a redefinition such that $S(g,a) = S(\tilde{g},a)$ where $\tilde{g} = f(g)$. The S-matrices will only agree for all values of $a$ if the beta functions satisfy $\beta_{\tilde{g}} =  \sum_g \beta_g  \partial_g \tilde{g}$.  In general, these new beta functions are not those arising in the scheme used before the field redefinition \cite{Latorre:2000qc, Einhorn:2001kj}.  

From the perspective of the correlation functions, such a change in scheme may be natural.  While the S-matrix is unchanged by the field redefinition, the correlation functions are not.  Our scheme keeps the correlation functions of the original description fixed.  However, the correlation functions of the theory after the field redefinition are not identical to those of the original, and thus we have no guarantee that the schemes should be simply related.



A similar argument for model-independent boundary conditions involves focusing on the F components of the weakly coupled fields to simplify the analysis.  Consider a model with hidden sector coupled to weakly coupled fields $A$ and $Q$ by the K\"{a}hler potential
\beq
\int d^4x d^4 \theta \left[ A^{\dag}A + (a X^{\dag} Q^{\dag} A + h.c.) + b X^{\dag}X Q^{\dag}Q \right].
\eeq
We want to integrate out the F component of A ($F_A$) and focus on the running of $F_{Q}$.  Because of the form of the K\"{a}hler potential, the only term involving $F_Q$ is
\beq
\int d^4 x \lambda F_Q F^{\dag}_Q \phi_X^{\dag} \phi_X,
\eeq
where $\phi_X$ is the lowest component of $X$ and $\lambda$ is some coupling.  Because nothing else couples to $F_Q$ we expect that $\lambda$ runs to zero with dimension $\Delta_{X^{\dag}X}$.  If we integrated out $F_A$ when $X$ is weakly coupled, we would find $\lambda = b-|a|^2$ and we might again conclude that this combination is forced to zero.  To see why this is incorrect, let us consider integrating out $F_A$ at strong coupling.

Integrating out $F_A$ should give all the terms required by supersymmetry.  Since supersymmetry should be maintained at all scales, we should get the same answer no matter when we choose to eliminate it.  If we try to integrate out $F_A$ at strong coupling, we will generate a term of the form
\beq
\int d^4 x \phi_X^{\dag}\phi_X F_Q F^{\dag}_Q(x) \left(b - |a|^2\int d^4y \delta^4 (x-y) C |x-y|^{\Delta_{X^{\dag}X}-2\Delta_X} \right),
\eeq
where we have used $F(x) F(y) \sim \delta^4(x-y)$ for a weakly coupled field.  When ${\Delta_{X^{\dag}X}-2\Delta_X} = 0$ this gives a nice result for $\lambda$, but more generally it is some singular function.  It is not clear how one should even define this function, as it quite badly behaved.  Furthermore, this term must be defined carefully in order for supersymmetry to be maintained.

Fortunately, our analysis from Section \ref{section:beta} tells us exactly how to deal with this divergence.\footnote{In the case, $\Delta_{X^{\dag}X}>2\Delta_X$ this term in the OPE is a zero rather than a pole.  However, to regulate the theory one must cutoff all integrals at $|x-y| \sim a$ like it is a divergent term.} In fact, the above term is already included in (\ref{OPE}) in terms of components.  One can even check that when ${\Delta_{X^{\dag}X}-2\Delta_X} = 0$, the divergent term in the OPE is just the delta function (as it should be for the $F$ components) and we get no additional contribution to the beta function.  Therefore, our beta function calculation gives us the correct definition of this term required to maintain manifest supersymmetry.  Working with components suggests that some combination of couplings runs to zero under RG flow.  However, if the flow is not restricted to a single weakly coupled fixed point, care is required to determine the correct linear combination of the couplings.  As we have seen from the OPE, the correct linear combination is a function of scale in the presence of non-trivial RG flow.

\section{Discussion and Conclusion}

In this paper, we have determined the form of the beta functions for soft parameters in the MSSM due to hidden sector renormalization using conformal perturbation theory.  These techniques work to all order in the hidden sector parameters, although they do not include subleading contributions due to Standard Model couplings.  These results are quite general, and should be of use in many contexts where the hidden sector is approximately conformal and strongly coupled.

The primary motivation for this work was the suggestion that strong dynamics can solve the $\mu$ problem.  These models are quite elegant in that they can solve the $\mu$ problem without fine tuning or baroque messenger sectors.  However, previously proposed `universal' boundary conditions, independent of the detailed hidden sector, would have severely constrained such models by the requirements of electroweak symmetry breaking.  Our analysis has shown that, in fact, the boundary conditions for soft terms arising from a strongly-coupled hidden sector depend explicitly on the details of the strongly-coupled theory. As a result, the model-dependence in the high-scale boundary conditions resurrects the prospects of hidden sectors that naturally solve the $\mu$ problem and are consistent with current data.

Although our results for hidden sector renormalization are rather general, concrete examples are required for further progress. To date, there are no known $\mathcal{N} = 1$ 4d SCFTs exhibiting the desired relationship $2 \Delta_X < \Delta_{X^\dag X},$ though such a relation is allowed in principle; it would be interesting to determine whether such theories exist. A concrete SCFT would illuminate many of the issues discussed in our analysis. Such examples would also allow a more precise determination of model-dependent results such as the OPE coefficient $C,$ which will partly determine the soft parameters of the Higgs sector. A concrete example is likewise required to elucidate the details of conformal symmetry breaking, which determines the exact relationship between hidden-visible couplings and SUSY-breaking soft masses. That being said, we do not expect these details to substantially alter our analysis, which should be of general relevance to theories with strongly coupled hidden sectors.  

\acknowledgments
 We are gratefully indebted to Martin Schmaltz for numerous helpful discussions and probing questions that motivated many of the results in this paper.  We would like to thank Savas Dimopoulos, Michael Dine, David Shih, Stuart Raby and Mithat Unsal for helpful discussions.  We are especially grateful to Steve Shenker for several extremely enlightening conversations on the subject of the renormalization group.  DG is supported in part by NSERC, the Mellam Family Foundation, the DOE under contract DE-AC03-76SF00515 and the NSF under contract PHY-9870115. NJC is supported in part by the NSF GRFP, the NSF under contract PHY-9870115, and the Stanford Institute for Theoretical Physics.

\bibliography{rg}
\end{document}